\documentclass[usenatbib,usegraphicx]{mn2e}
\usepackage{times}
\usepackage[intlimits]{amsmath}

   \title[Analytical approximation for rotating barotropes]{Analytical approximation for the structure of
   differentially rotating barotropes}

   	\author[A.~Odrzywo\l{}ek]{A.~Odrzywo\l{}ek\thanks{
	E-mail: odrzywolek@th.if.uj.edu.pl}
			\\
	Institute of Physics, Jagiellonian University, Reymonta 4, 30-059 Krakow, Poland}

\date{Released 2003 May 18}

\pagerange{\pageref{firstpage}--\pageref{lastpage}} \pubyear{2003}
\begin{document}
\label{firstpage}

\maketitle

\begin{abstract}
Approximate analytical formula for density distribution in differentially
rotating stars is derived. Any barotropic EOS and conservative rotation law
can be handled with use of this method for wide range of differential rotation
strength. Results are in good qualitative agreement
with comparison to the other methods. Some applications are suggested
and possible improvements of the formula are discussed.
\end{abstract}


\section{Introduction}

Theory of self-gravitating rotating bodies seems to be an
unlimited reservoir of difficult problems hardly tractable
even under severe simplifications. It is a subject  of scientific effort since
1742 when Maclaurin has initiated this field by his studies on
incompressible rotating ellipsoids \citep{maclaurin}. Development of modern numerical
calculations resulted in progress in practical applications
nowadays such as e.g. 3D hydrodynamical simulations of rotation of complex
objects.

Analytical approach has succeeded for constant density, incompressible
bodies. Work of Maclaurin, Jacobi, Poincare,
Schwarzschild and many others has explained the behaviour of those objects almost
completely \citep{lyttleton}.
Behaviour of slowly rotating polytropes  has  been
calculated by \citet{chandra}. By applying the differential
equation of hydrostatical equilibrium modified by rotation
he reduced the problem to an ordinary differential equation.
This method however works only for a uniform rotation.
This list would be incomplete without the Roche model. 
It's simplicity makes it a very powerful tool for
understanding behaviour of rotating objects.
Present computational methods allow one to
handled numerically two- and three-dimensional
problems with complicated governing equations

In this paper we present simple analytical
approach which can treat differentially rotating
compressible barotropic stars in case of slow or moderately fast
rotation. This model could fill a gap between simple analytical methods used
for e.g. Maclaurin spheroids or Roche model, and complicated numerical methods
such as e.g. HSCF  \citep{hachisu}, or those applying straightforward
Newton-Raphson   technique \citep{eriguchi}.

\section{Formulation of the problem}

We attempt to find a density distribution (iso-density contours)
of a~single self-gravitating object under the following assumptions:
\begin{enumerate}
\item{Barotropic EOS} $p=p(\rho)$
\item{Simple rotation $\mathbf{v} = r \, \Omega \, \mathbf{e}_{\phi}$ with
angular velocity dependent only on the distance from rotation axis
\mbox{$\Omega = \Omega(r)$}}
\item{ Newtonian gravity}
\item{ Axisymmetric density distribution}
\item{ We seek solutions for stationary objects in full mechanical equilibrium,
i.e. all quantities are time-independent}
\end{enumerate}

With properties (i)--(v) satisfied, the
Euler equation becomes, in cylindrical coordinates ($r$, $z$, $\phi$):
\begin{equation}
\label{euler}
r\,\Omega(r,z)^2\, \mathbf{e}_r = \frac{1}{\rho} \nabla p + \nabla \Phi_g
\end{equation}
Continuity equation is then fulfilled automatically.
Introducing centrifugal potential:
\begin{equation}
\label{centrifugal}
\Phi_c(r) = - \int_0^r \tilde{r}\;  \Omega(\tilde{r})^2\; d\tilde{r}
\end{equation}
and enthalpy:
\begin{equation}
\label{enthalpy}
h(\rho) = \int \frac{1}{\rho} dp
\end{equation}
we get a simple equation:
\begin{equation}
\nabla [ h(\rho) + \Phi_g + \Phi_c ] = 0
\end{equation}
with a solution
\begin{equation}
\label{main_eq}
h(\rho) + \Phi_g + \Phi_c = C = const
\end{equation}

Equation (\ref{main_eq}) is the most important equation  in the study of the
structure of  rotating
stars under conditions (i)--(v). We define the integration constant
in~eq.~(\ref{enthalpy}) to
be such that the enthalpy  satisfies the condition $h(\rho=0)=0$.
 The only term which we haven't specified  yet is the gravitational potential $\Phi_g$.
If we use the Poisson
equation:
\begin{equation}
\Delta \Phi_g = 4\, \pi\, G \rho
\end{equation}
where $G$ is the gravitational constant, the equation~(\ref{main_eq}) becomes
a~non-linear second-order
differential equation. This form, however, is very inconvenient, because we
have to
specify boundary conditions at a surface of the
star,\footnote{The fact of the
surface of non-rotating stars is spherical allows us to specify boundary
conditions with
one real number -- the radius. Generally, in rotating objects we have a surface
represented by some function of two variables which has to be determined.}
which is unknown {\it a priori}.
More powerful is an integral form of eq.~(\ref{main_eq}) obtained by
substitution:
\begin{equation}
\label{int_potential}
\Phi_g(\mathbf{r}) = -G  \int
\frac{\rho(\tilde{\mathbf{r}})}{|\mathbf{r}-\tilde{\mathbf{r}}|}
d^3 \tilde{\mathbf{r}}
\end{equation}
This integral form has been used in very successful numerical algorithm
developed originally by  \cite{ostriker}, and recently
improved
by \cite{hachisu} and by \cite{eriguchi}.
This form will be also used to derive our approximation formula in the next
section.

\section{Approximation for the density distribution}

The {\em integral equation} form of eq.~(\ref{main_eq}) is:
\begin{equation}
\label{int_eq}
h(\rho)+\mathcal{R}(\rho) + \Phi_c = C
\end{equation}
where $\mathcal{R}$ is the integral operator acting on the density performing
the integration on the right-hand side of eq.~(\ref{int_potential}) over
entire volume
of the star. We define the surface of the star to be manifold consisting of
points where $\rho=0$, $h=0$.
Explicit form of the operator $\mathcal{R}$ in terms of coordinates will not
be needed.
For a given  EOS and for a fixed rotation law i.e. for given functions
$h(\rho)$
and $\Phi_c$, the only free parameter is the constant $C$. The values of $C$
label a
family of the stellar models with the same EOS and rotation pattern, which
differ
in total mass and maximum density\footnote{
In differentially rotating stars the central density may be, but generally is
not, the maximum
density.}  etc.

Eq.~(\ref{int_eq}) has a form of the Hammerstein non-linear integral equation
\citep{hammer} and can be rewritten in a canonical form:
\begin{equation}
\label{eq_canon}
f = \mathcal{R} \left [ F(f) \right ]
\end{equation}
where:
\begin{equation}
f = C - \Phi_c - h(\rho), \qquad F(f) = h^{-1}(f + \Phi_c - C)
\end{equation}
In case of linear function $F$, eq.~(\ref{eq_canon}) could be easily solved
by the von~Neumann series. This strongly suggests to try the following
iteration scheme:
\begin{eqnarray}
\label{seq}
\nonumber
f_1& = \mathcal{R}[ F(f_0)], \\
\nonumber
f_2& = \mathcal{R}[ F(f_1)], \\
&\cdots \\
\nonumber
f_n& = \mathcal{R}[ F(f_{n-1})] \\
\nonumber
&\cdots
\end{eqnarray}
Indeed, an iteration procedure of this type was successfully applied in the
so-called
self-consistent field method (\citealt{ostriker}, \citealt{hachisu}).
We have introduced the canonical form to ensure that
the first-order approximation is found in a correct order i.e. by using
the first line of the sequence~(\ref{seq}). When we go back to
non-canonical form (\ref{int_eq}) the first line of (\ref{seq}) takes the form:
\begin{equation}
\label{correct_order}
C- \Phi_c - h(\rho_1) = \mathcal{R}\left(\rho_0\right)
\end{equation}

From eq.~(\ref{correct_order}) above we can find the
first-order deviation from sphericity. It seems impossible
at first sight to avoid explicit integration in eq.~(\ref{correct_order}).
In case of a general $\rho_0$, this is true. But let us look
at  equation
(\ref{int_eq}) in case of vanishing centrifugal potential $\Phi_c$, i.e. with
no rotation:
\begin{equation}
\label{non_rot}
h(\rho)+\mathcal{R}(\rho) = C
\end{equation}
When we use a function which satisfies eq.~(\ref{non_rot})
as zero-order approximation:
\begin{equation}
\label{zero_deg}
h(\rho_0)+\mathcal{R}(\rho_0) = C_0
\end{equation}
integration in eq.~(\ref{correct_order}) can be easily eliminated:
\begin{equation}
C- \Phi_c - h(\rho_1) = \mathcal{R}(\rho_0) = C_0 - h(\rho_0)
\end{equation}
Finally, our formula takes the form:
\begin{equation}
h(\rho_1) = h(\rho_0) - \Phi_c + C - C_0
\end{equation}
or simpler, using the enthalpy  ($h(\rho_0) \equiv h_0$, $h(\rho_1) \equiv
h_1$):
\begin{equation}
\label{formula}
h_1 = h_0 - \Phi_c + C - C_0
\end{equation}
Functions used as zero-order approximation ($\rho_0$ or $h_0$)  are simply
density
and enthalpy distributions of non-rotating barotropic stars.
In case of polytropic EOS, $p=K \rho^{\gamma}$, these quantities are
given by Lane-Emden functions. In more general case we have
to find solution of the ordinary differential equation of the hydrostatic
equilibrium.

The only unanswered question is what are the value of constants $C$ and
$C_0$\footnote{
For a given EOS the function $\rho_0$ gives $C_0$ and {\it vice versa}.}.
It is an essential part of this work, so we decided to explain it in a
separate section.

\section{Adjusting constants}

When we try to find the enthalpy distribution using the formula (\ref{formula})
we have to find the best zero-order function $h_0$ and the value of $C_0$
given by $h_0$. An equivalent problem is to find the equation for which the
function
$h_0$ and the value $C_0$ are best zero-order approximations  -- in this case
we seek for $C$. We consider the latter case, as we
have to find only one real number.
Let us denote:
\begin{equation}
\label{delta}
\Delta C = C_0 - C.
\end{equation}
In our approximation, the equation for the first-order enthalpy distribution $h_1$,
in terms of the initial spherical
distribution $h_0$ and rotation law is, from (\ref{formula}):
\begin{equation}
\label{deltaC}
h_1 = h_0 - \Phi_c - \Delta C
\end{equation}
where $\Delta C$ is still to be determined.

\begin{figure}
\includegraphics[height=84mm,angle=270]{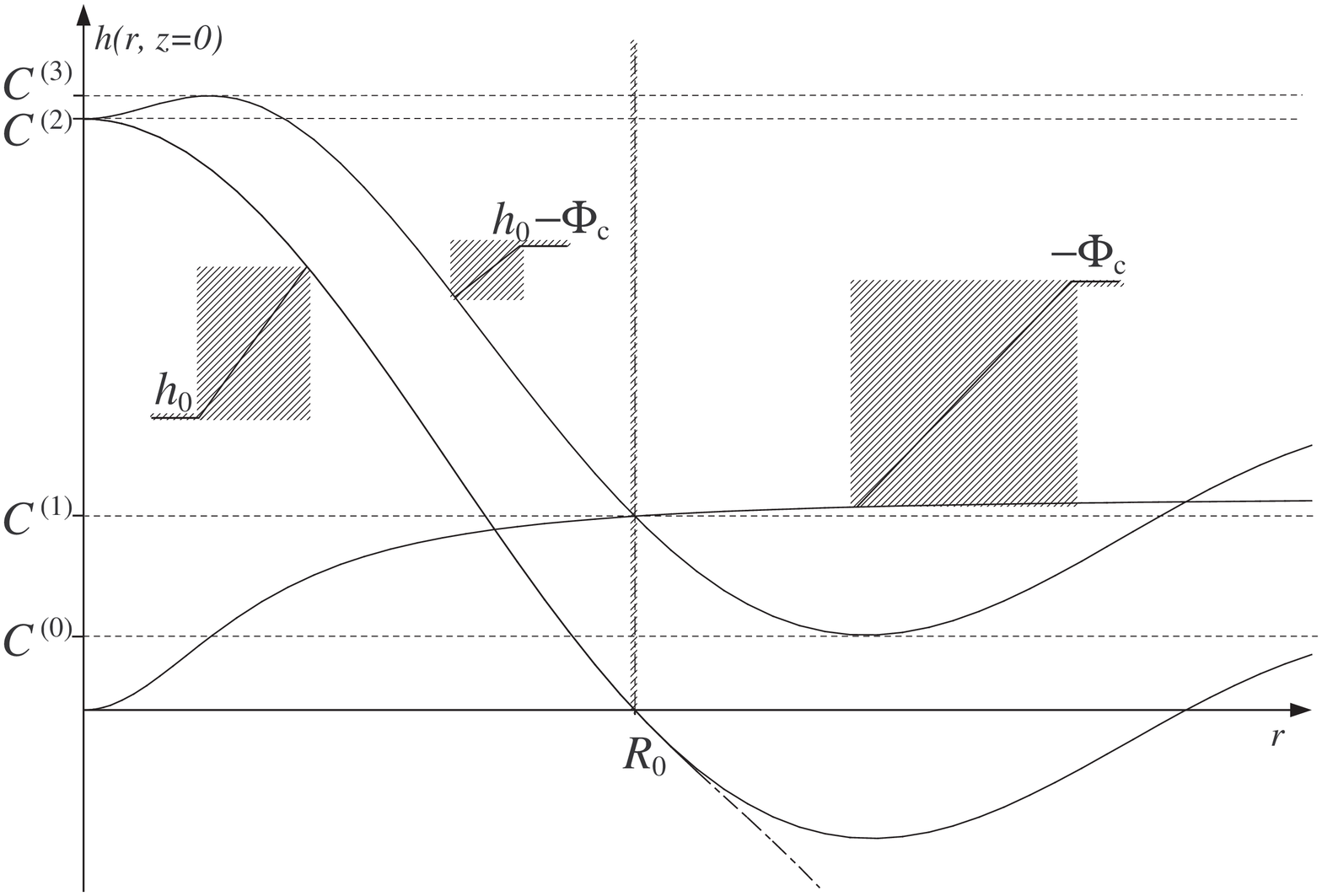}
\caption{Schematic graphs of the two first terms in
eq.~(\ref{deltaC}), explaining the meaning of $\Delta C$. The most general
case is shown. In some particular cases $C^{(0)}$ and $C^{(2)}$ may not exist -- this depends
on the both $h_0$ and $\Phi_c$. Vertical dot-dashed line shows that
$C^{(1)}=\Phi_c(R_0)$, where $R_0$ is radius of non-rotating star. The dashed curve
fragment below the axis reflects ambiguity of the Lane-Emden function continuation
to negative values, as described in the text. This figure prepared with use of
eq.~(\ref{simple_case}) for $z=0$.}
\label{terms}
\end{figure}

The terms in eq.~(\ref{deltaC}) behave as follows:
\begin{enumerate}
\item{`$h_0$' is spherical enthalpy distribution, thus it is only a function
of the radius, has a maximum at the centre, and goes monotonically to
zero, where usually is cut. However, from mathematical point of view, Lane-Emden
functions extend beyond the first zero point with negative function values.}
\item{`$-\Phi_c$' is a monotonically increasing function of distance from the
rotation axis. It starts with zero at the rotation axis. It does not change
the enthalpy along the axis of symmetry. The strongest enthalpy increase takes
place
along the equatorial plane. }
\item{`$\Delta C$' shifts the sum of positive functions $h_0$ and $-\Phi_c$
down.}
\end{enumerate}
At first sight, shifting down by $\Delta C$  seems
not needed (i.e. one would adopt $\Delta C =0 $ ) , because we
obtain correct qualitative behaviour -- the star is expanded along equator.
But often it is enough to introduce slow rotation to get positive value of
$h_0-\Phi_c$ (i.e. for our approximation to the enthalpy in this case) for any $r>0,z=0$ i.e. equatorial
radius becomes infinite.
It leads directly to physically unacceptable results -- infinite volume and mass.
So the value $\Delta C$ plays a non-trivial role and has to be found.
Fig.~\ref{terms} shows the behaviour  of  all terms in eq.~(\ref{deltaC}) along the equator
of the star, where the rotation acts most strongly. Horizontal
lines show points where enthalpy is cut for a given value of $\Delta C$.

We can distinguish some important values:
\begin{enumerate}
\item{For $\Delta C <C^{(0)}$ we obtain infinite radius of a star.
These
values obviously have to be rejected.}
\item{For $C^{(0)} < \Delta C < C^{(1)}=\Phi_c(R_0)$, where
$R_0$ is the
radius of a zero-order density distribution, we get finite volume of a star,
but
we use extension of $h_0$ with negative values. This introduces some problems
which we discuss later in the article, although the resulting enthalpy and
density are positive and physically acceptable.}
\item{For $C^{(1)} < \Delta C < C^{(2)}=h_0(r=0,z=0)$ we get
a density distribution which is topologically equivalent to the ball.}
\item{For $C^{(2)}<\Delta C < C^{(3)}=(h_0 - \Phi_c)_{max}$ we
get toroidal
density distribution.
This case exists only if strong differential
rotation is present. }
\item{For  $\Delta C > C^{(3)}$ the star disappears.}
\end{enumerate}
We expect to find the solution in the range $C^{(0)} < \Delta C <
C^{(2)}$
because we are looking for finite-volume non-toroidal stars.

One can try to find $\Delta C$ both analytically and numerically. To keep
the algebraic form and the simplicity of the formula, we now concentrate on the
former method.

When we substitute the formula (\ref{deltaC}) into our basic
equation (\ref{main_eq}) we get:
\begin{equation}
h_0-\Phi_c+C-C_0 + \Phi_g(h_1) + \Phi_c = C
\end{equation}
In this formula  we have made use of (\ref{delta}). After obvious
simplifications, using (\ref{zero_deg}) and denoting
$\mathcal{R}(\rho_0) = \Phi_g(h_0)$ we have:
\begin{equation}
\Phi_g(h_1) = \Phi_g(h_0)
\end{equation}
This equality is true only if $\rho_0 = \rho_1$. The same holds
for the enthalpy:
\begin{equation}
h_1=h_0.
\end{equation}
Using formula (\ref{deltaC}) again we finally obtain:
\begin{equation}
\label{deltaC1}
\Delta C = -\Phi_c
\end{equation}
Left-hand side of eq.~(\ref{deltaC1}) is constant, while the right-hand side
is a function of distance from the rotation axis, monotonically decreasing
from zero.
This equality holds only in trivial case $\Delta C = 0$ and $\Phi_c = 0$
with no rotation at all. In any other case (\ref{deltaC1}) cannot be fulfilled.
So instead we try another possibility and require that
\begin{equation}
\label{deltaC2}
\Delta C = - \widehat{\Phi_c}
\end{equation}
where `hat' denotes some mean value of the function $\Phi_c$.
We have chosen
\begin{equation}
\label{meanC}
\Delta C = - \widehat{\Phi_c} =
- \left ( \frac{4}{3} \pi {R_0}^3 \right )^{-1} \int_{V_0} \Phi_c \, d^3
\mathbf{r}
\end{equation}
Integration is taken over the entire volume $V_0 = \frac{4}{3} \pi R_0^3$ of a non-rotating initial
star with the radius $R_0$.
This choice of $\Delta C$ gives good results.
But using the mean value theorem:
\begin{equation}
\int_{V_0} \Phi_c \, d^3 \mathbf{r} = V_0 \, \widehat{\Phi_c},
\end{equation}
where $\widehat{\Phi}_c$ is some value of $\Phi_c$ in the integration area,
and taking in account monotonicity of the centrifugal potential $0 < -\Phi_c <
-\Phi_c(R_0) $
we get:
\begin{equation}
\Delta C = -\widehat{\Phi_c} < -\Phi_c(R_0)
\end{equation}
i.e. the value of $\Delta C$ is in the range $\Delta C < C^{(1)}$
from Fig.~\ref{terms}. It forces us to use negative values of non-rotating enthalpy.
Moreover, in case of polytropic EOS with fractional polytropic index\footnote{
Physically interesting cases like degenerate electron gas in non-relativistic case has fractional polytropic index $n=3/2$.}
Lane-Emden equation \citep{kippenhahn}:
\begin{equation}
\label{l-e}
\frac{1}{z^2} \frac{d}{dz} \left( z^2 \frac{dw}{dz} \right ) + w^n = 0
\end{equation}
has no real negative values, because of fractional power of negative term  $w^n$.
But we can easily write equation, with solution identically equal to solution of Lane-Emden equation for
$w>0$, and real solution for $w<0$ e.g.:
\begin{equation}
\label{abs}
\frac{1}{z^2} \frac{d}{dz} \left( z^2 \frac{dw}{dz} \right ) + |w|^n = 0
\end{equation}
But, for example, solution of the following equation:
\begin{equation}
\label{sign}
\frac{1}{z^2} \frac{d}{dz} \left( z^2 \frac{dw}{dz} \right ) +  \frac{|w|^{n+1}}{w} = 0
\end{equation}
is again identically equal to solution of Lane-Emden equation for $w>0$, but differs from eq.~(\ref{abs}) for $w<0$.
Fortunately, difference between solution of eq.~(\ref{abs})  (Fig.~\ref{terms}, below axis, dot-dashed) and
eq.~(\ref{sign}) (Fig.~\ref{terms}, solid) for $w<0$ is small if $|w| \ll 1$. Example from Fig.~\ref{terms} (for $n=1$) is
representative for other values of
$n$. We will use form (\ref{abs}) instead of the original Lane-Emden equation (\ref{l-e}) for calculations in this article.

To avoid problems with negative enthalpy we can put simply:
\begin{equation}
\label{deltaC3}
\Delta C = -\Phi_c(R_0)
\end{equation}
which is strictly boundary value $C^{(1)}$ from Fig.~\ref{terms}.
The great advantage of the eq.~(\ref{deltaC3}) is the possibility
to analytically perform the  integration of the centrifugal potential
(\ref{centrifugal}) for
most often used forms of $\Omega(r)$.  In contrast, in formula
(\ref{meanC}),
not only  angular velocity profile (\ref{centrifugal}), but also the centrifugal potential have
to be analytically integrable function.
In both cases (\ref{meanC}, \ref{deltaC3}) however, possibility of analytical
integration
depends on the form of $\Omega(r)$.
The value of $\Delta C$ from eq.~(\ref{deltaC3}) also gives reasonable iso-density contours,
cf. Fig.~\ref{v-const} and \ref{j-const}, but global accuracy is poor (Table \ref{A02jconstMS}).

As we noticed, the best value of $\Delta C$ in formula~(\ref{deltaC})
could be found numerically. For example, we can use virial theorem formula for rotating stars (cf. \citealt{tassoul}):
\begin{equation}
\label{virial}
2\, E_k - \left| E_g \right | + 3\, \int p\, d^3 \mathbf{r} = 0,
\end{equation}
where $E_k$ and $E_g$ is the rotational kinetic energy and the gravitational energy, respectively.
We define, so-called virial test parameter $Z$:
\begin{equation}
\label{z}
Z =\frac{2\, E_k - \left| E_g \right | + 3(\gamma-1)U}{|E_g|}
\end{equation}
where we introduced internal energy:
\begin{equation}
U = \frac{1}{\gamma-1} \int p\, d^3 \mathbf{r}
\end{equation}
Parameter Z is very common test of the global accuracy for rotating stars models.
We may request that our enthalpy satisfy (\ref{virial}), i.e. we choose $\Delta C$ from equation:
\begin{equation}
\label{deltaVT}
Z \left ( h_0 - \Phi_c - \Delta C \right ) = 0
\end{equation}
We can  find $\Delta C$ from equation eq.~(\ref{deltaVT}) numerically only.
\begin{figure}
\includegraphics[width=84mm]{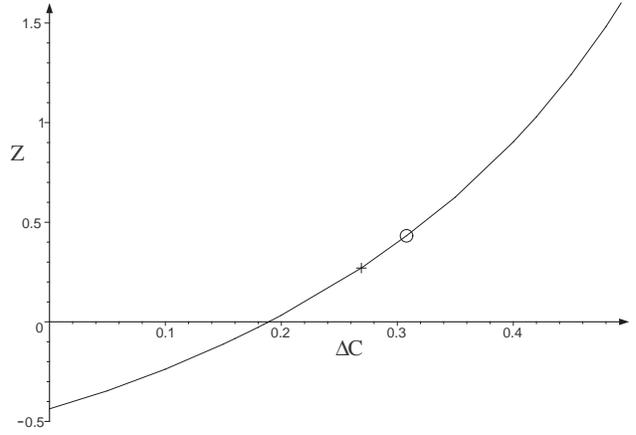}
\caption{Behaviour of virial test parameter $Z$ vs $\Delta C$ for  $n=3/2$ polytropic model (\ref{n32_case})
with $j$-const angular velocity profile for $\Omega_0 = 1.5$ and $A=0.2 R_0$. Proper choose of $\Delta C$
can give density distribution
satisfying virial theorem to arbitrary accuracy. $\Delta C\!=\!0.308$ from eq.~(\ref{deltaC3}) is represented by circle.
Cross marks value $\Delta C\!=\!0.269$
given by formula (\ref{meanC}). Virial theorem will be satisfied when we take $\Delta C$, given by
intersection of Z with horizontal axis i.e. $\Delta C\! =\! 0.189$.}
\label{VT_vs_deltaC}
\end{figure}

As it is shown on Fig.~\ref{VT_vs_deltaC},
we can find approximation of the rotating polytrope structure in form (\ref{deltaC})
satisfying virial theorem (\ref{virial}) up to accuracy limited only by numerical precision.

Values of $\Delta C$ obtained with (\ref{deltaC3}), (\ref{meanC}) and from virial test (\ref{deltaVT})
are compared on Fig.~\ref{3xDelta}. Some of the global model properties are very sensitive
to value of $\Delta C$ (cf. Figs \ref{VT_vs_deltaC} and \ref{Etot_vs_j2}).

Because virial test is unable to check accuracy of our model,
we may also try to compare directly eq.~(\ref{deltaC}) with enthalpy distribution
from the numerical calculations $h_{\mathrm{num}}$ of e.g. \cite{hachisu}, \cite{eriguchi}
and find $\Delta C$ minimizing e.g. the following formula:
\begin{equation}
\int \left[ h_1(\Delta C) - h_{\mathrm{num}} \right ]^2 d^3\mathbf{r}= \mathrm{min}
\end{equation}
This method however, requires numerical results (e.g. enthalpy distribution) in machine-readable form.

\section{Approximate formula accuracy}

In the above sections, we tried to be as general as possible. Now we
give some examples, and test accuracy of approximation.

In case of polytropic EOS $p=K \rho^{\gamma}$ the enthalpy is:
\begin{equation}
h(\rho) = \frac{K \gamma }{\gamma-1} \rho^{\gamma-1}
\end{equation}
Zero-order approximation of density (density of non-rotating polytrope, \citealt{kippenhahn}) 
with n-th\footnote{$\gamma = 1 + \frac{1}{n}$}
Lane-Emden function $w_n$ is:
\begin{equation}
\rho_0 = \rho_c \left [ w_n(A r)   \right ]^n, \qquad A^2 = \frac{4 \pi G}{n K \gamma } \rho_c^{\frac{n-1}{n}}
\end{equation}
and our formula for density becomes:
\begin{equation}
\rho_1 = \left[ {\rho_c}^{1/n} w_n - \frac{1}{n K \gamma}(\Phi_c+\Delta C)
\right ]^n
\label{polytropic_case}
\end{equation}
where $\Delta C$ is calculated from (\ref{meanC}),  (\ref{deltaC3}) or (\ref{deltaVT}).

In certain cases Lane-Emden functions are elementary functions as e.g. $w_1$.
In cases like this our formula may be expressed even by elementary functions.
For example, for $n\!=\!1$,
$K\!=\!1/2$,
$4 \pi G\!=\!1$,
$\rho_c\!=\!1$,
$\Omega(r)\!=\!{\Omega_0}/{(1+r^2/A^2)}$ and $\Delta C$
from eq.~(\ref{deltaC3}) we get a simple formula:
\begin{equation}
\rho_1(r,z) = \frac{\sin \sqrt{r^2+z^2}}{\sqrt{r^2+z^2}}+
\frac{1}{2} \frac{\Omega_0^2 A^2 r^2}{1+r^2/A^2}
-\frac{1}{2} \frac{\Omega_0^2 A^2 \pi^2}{1+\pi^2/A^2}
\label{simple_case}
\end{equation}

Functions like this can easily be visualized on a 2D plot. Figure \ref{terms}
has been made from the formula (\ref{simple_case}) while figures \ref{j-const}
and \ref{v-const} from eq.~(\ref{n32_case}).

Now we concentrate on $n=3/2$ polytrope. In our calculations and figures we will use $4 \pi G = 1$,
$\rho_c=1$ and $K=2/5$. Now formula (\ref{polytropic_case}) becomes:
\begin{equation}
\rho_1 = \left( w_n - \Phi_c-\Delta C \right )^{3/2}
\label{n32_case}
\end{equation}
Iso-density contours of $\rho_1$ from (\ref{n32_case}) are presented on Fig.~\ref{v-const} and Fig.~\ref{j-const}.

To test accuracy of approximation we have calculated axis ratio, total energy, kinetic to gravitational energy ratio,
and dimensionless angular momentum.
Axis ratio is defined as usual as:
\begin{equation}
\label{axisratio}
\mathrm{Axis\; Ratio} = \frac{R_z}{R_{eq}}
\end{equation}
where $R_z$ is distance from centre to pole and $R_{eq}$ is equatorial radius.
Total energy $E_{tot}$:
\begin{equation}
\label{Etot}
E_{tot} = (E_k+E_g+U)/E_0
\end{equation}
is normalized by:
\begin{equation}
E_0 = (4 \pi G)^2   \frac{M^5}{J^2}
\end{equation}
and dimensionless angular momentum is defined as:
\begin{equation}
\label{j_squared}
j^2  =  \frac{1}{4 \pi G}   \frac{J^2}{M^{10/3}} \, \rho_{max}^{1/3}
\end{equation}
where $M$ and $J$ are total mass and angular momentum, respectively; $\rho_{max}$ is maximum density.
Quantities (\ref{axisratio})-(\ref{j_squared}) are computed numerically from (\ref{n32_case}),
with given angular velocity $\Omega(r)$ and chosen $\Delta C$.

\subsection{Influence of $\Delta C$}
{

\begin{figure}
\includegraphics[width=84mm]{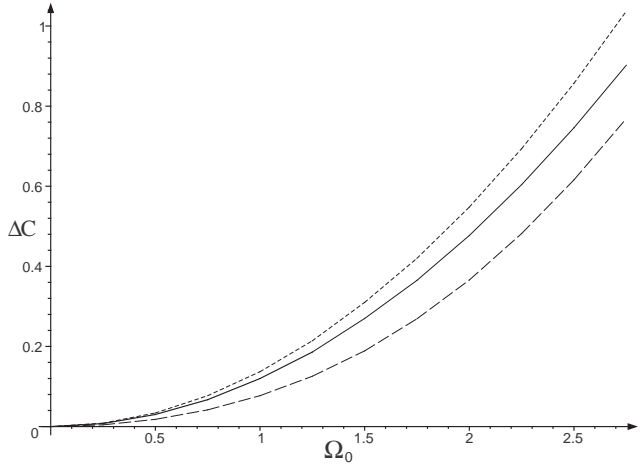}
\caption{Dependence of $\Delta C$ on $\Omega_0$, given by eq.~(\ref{deltaC3}) (dotted),
calculated from (\ref{meanC}) (solid)
and given by virial theorem constrain (\ref{deltaVT}) (dashed).
Both values estimated
by (\ref{meanC}, \ref{deltaVT}) are below $\Delta C \! = \! -\Phi_c(R_0)$ i.e. $C^{(1)}$ from Fig.~\ref{terms}.
It shows, that continuation of Lane-Emden
equation to negative values is required for successful approximation of the rotating body structure.
Density distribution was given by eq.~(\ref{n32_case}) with $j$-const angular velocity with $A=0.2 R_0$.
}
\label{3xDelta}
\end{figure}

\begin{table*}
\begin{tabular}{ccrlrcccc}
\hline
$\Omega_0$&Axis~Ratio&$j^2 \qquad$& $\frac{E_k}{|E_g|}$  & $\frac{E_g+E_k+U}{E_0}$ & Virial test  $Z$&$\widetilde{Z}$& $\Delta C$ & $\Delta \widetilde{C}$\\
\hline \\
$0.25  $&$ 1.01$&$  8.41 \times 	10^{-5} $&$0.004 $&$ -2.50 \times   10^{-6}  $&$ 0.01 $&$0.01$&$ 0.01$&$0.01$\\
$0.50  $&$ 1.05$&$  3.38 \times 	10^{-4} $&$0.02  $&$ -1.00 \times   10^{-5}  $&$ 0.03  $&$0.04$&$ 0.03$&$0.03$\\
$0.75  $&$ 1.15$&$  7.64 \times 	10^{-4} $&$0.04  $&$ -2.20 \times   10^{-5}  $&$ 0.07  $&$0.10$&$ 0.07$&$0.08$\\
$1.00  $&$ 1.19$&$  1.37 \times 	10^{-3} $&$0.07  $&$ -3.80 \times   10^{-5}  $&$ 0.12  $&$0.17$&$ 0.12$&$0.14$\\
$1.25  $&$ 1.31$&$  2.17\times 	10^{-3} $&$0.10  $&$ -5.56 \times 10^{-5}  $&$ 0.19  $&$0.28$&$ 0.19$&$0.21$\\
$1.50  $&$ 1.46$&$  3.12 \times 	10^{-3} $&$0.15  $&$ -7.38 \times 10^{-5}  $&$ 0.27  $&$0.43$&$ 0.27$&$0.31$\\
$1.75  $&$ 1.68$&$  4.40 \times 	10^{-3} $&$0.20  $&$ - 8.93\times 10^{-5}  $&$ 0.36  $&$0.63$&$ 0.37$&$0.42$\\
$2.00  $&$ 1.98$&$  5.90 \times 	10^{-3} $&$0.26   $&$ - 9.89\times 10^{-5}  $&$ 0.47  $&$0.89$&$ 0.48 $&$0.55$\\
$2.25  $&$ 2.45$&$  7.55\times 	10^{-3} $&$0.32   $&$ - 1.01\times 10^{-4}  $&$ 0.57    $&$1.22$&$ 0.60 $&$0.69$\\
$2.50  $&$ 3.30$&$  9.30 \times 	10^{-3} $&$0.38   $&$ - 0.94\times 10^{-4}  $&$ 0.67    $&$1.66$&$ 0.75 $&$0.86$\\
$2.75  $&$ 5.80$&$  1.10 \times 	10^{-2} $&$0.44   $&$ - 0.84\times 10^{-4}  $&$ 0.75    $&$2.22$&$ 0.90 $&$1.04$\\
\hline
\end{tabular}
\caption{Properties of $n=3/2$ polytropic model (\ref{n32_case}) with $j$-const rotation law and $A=0.2 R_0$.
$\Delta C$ is calculated from eq.~(\ref{meanC}) and $\Delta \widetilde{C}$ from eq.~(\ref{deltaC3}). Virial
test in the latter case is labeled by $\widetilde{Z}$. By little change from $\Delta \widetilde{C}$ to $\Delta C$
one can notice significant improvement of virial test. In both cases virial test suggest strong deviation
from equilibrium, especially for strong rotation.
Actually, virial test is very sensitive to $\Delta C$, cf.~Fig.~\ref{VT_vs_deltaC}.
We can require virial theorem to be satisfied, by use of another value for $\Delta C$, solution of
eq.~(\ref{deltaVT}). Results can be improved significantly this way --
compare with Table~\ref{j-const02} and Fig.~\ref{Etot_vs_j2}.
}
\label{A02jconstMS}
\end{table*}

\begin{figure}
\includegraphics[width=84mm]{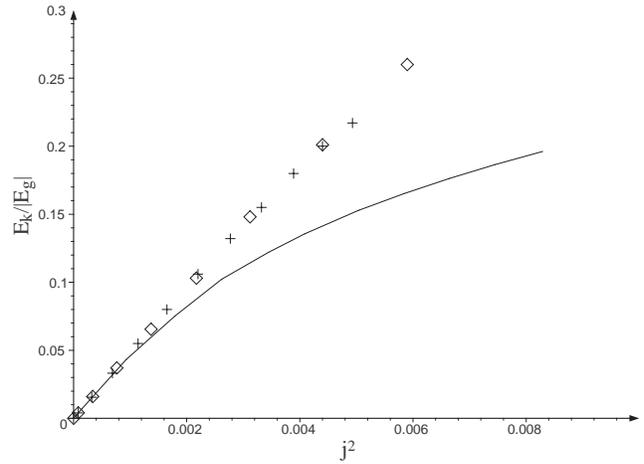}
\caption{
$E_k/|E_{g}|$ ratio as a function of the square of dimensionless angular momentum $j^2$ for
our model (\ref{n32_case}) with $n=3/2$, $\Omega_0 = 1.5$ and $A=0.2 R_0$.
Solid line represent numerical results of \citep{eriguchi}.
We see that our formula behaviour is in good agreement with
numerical results if $E_k/|E_{g}| \ll 0.1$. Results using $\Delta C$
from eq.~(\ref{meanC})  are marked by crosses. Results satisfying virial theorem ($\Delta C$ from eq.~(\ref{deltaVT}))
are represented by diamonds. As it is apparent from figure above, $\Delta C$ has no influence on this relation,
and can't improve accuracy of the formula (\ref{formula}).
}
\label{TWvsj2}
\end{figure}

\begin{figure}
\includegraphics[width=84mm]{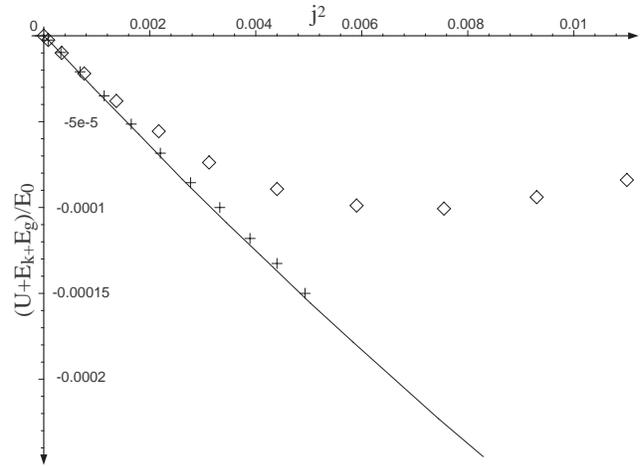}
\caption{Total energy versus $j^2$. We can see great improvement of results with $\Delta C$ from
eq.~(\ref{deltaVT}) marked by `$+$'. $\Delta C$ from eq.~(\ref{meanC}) ($\diamond$) gives  incorrect behaviour
of the total energy.  Solid line represent results of \citet{eriguchi}, cross and diamonds are result derived from
our approximate formula (\ref{n32_case}) with $\Delta C$ from virial test (\ref{deltaVT}) and eq.~(\ref{meanC}),
respectively. Approximation satisfying virial equation gives results resembling numerical calculations.
Parameters of model are given in caption of Fig.~\ref{TWvsj2}.
}
\label{Etot_vs_j2}
\end{figure}

We have made detailed comparison of our $n=3/2$ model (\ref{n32_case}) with $j$-const rotation law
and $A=0.2 R_0$ (middle row of Fig.~\ref{j-const}) for different values of $\Delta C$
with results of (\citealt{eriguchi}, Table 1b).
Table \ref{A02jconstMS} show our results for $\Delta C$ from eq.~(\ref{meanC}). Value of
$\Delta C$ from eq.~(\ref{deltaC3}) and corresponding virial test parameter $Z$ is included here
for comparison. Table \ref{j-const02} shows global properties of our approximation with
$\Delta C$ equal to the solution of eq.~(\ref{deltaVT}), i.e. satisfying virial theorem.

Direct comparison of values from Table~\ref{A02jconstMS} and table Table~\ref{j-const02}
to Table~1b of \citet{eriguchi} may be difficult, because our driving parameter is central angular velocity
$\Omega_0$, while \citet{eriguchi}, following successful approach of \citet{hachisu}, use axis ratio (\ref{axisratio}).
More convenient in this case is comparison of figures prepared from data found in Table 1b of
\citet{eriguchi} and our tables. This is especially true, because axis ratio isn't well predicted by our formula
(cf. Fig.~\ref{axisratio_vs_TW} and Fig.~\ref{axisratio_vs_j2}),
while global properties ($E_{tot}$, $j^2$, $E_k/|E_g|$, cf. Fig.~\ref{TWvsj2},~\ref{Etot_vs_j2})
and virial test $Z$ are in good agreement if $E_k/|E_g| \ll 0.1$.

Fig.~\ref{TWvsj2} shows that our approximation is valid until
$E_k/|E_g| \simeq 0.05$, and begins to diverge
from numerical results strongly for $E_k/|E_g| \geq 0.1$. Both values of $\Delta C$ (\ref{meanC},\ref{deltaVT})
give similar behavior here. However, $\Delta C$ from virial test produces better results, and $E_k/|E_g|$ values
are more sensible for the strongest rotation.

In contrast, total energy (\ref{Etot}) is very sensitive to $\Delta C$. Value of $\Delta C$ from eq.~(\ref{meanC})
produces wrong result. $E_{tot}$ begin to increase for $j^2 \geq 0.007$, while numerical results give
monotonically decreasing $E_{tot}$. Use of $\Delta C$ from (\ref{deltaVT}) instead, gives correct result,
cf. Fig.~\ref{Etot_vs_j2}.

While global properties of our model are in good agreement with numerical results for $E_k/|E_g| \ll0.1$,
axis ratio tends to be underestimated, even for small values of $j^2$. Fig.~\ref{axisratio_vs_TW}
and Fig.~\ref{axisratio_vs_j2} show minor improvements when we use $\Delta C$ from
virial test (\ref{deltaVT}) instead of mean value (\ref{meanC}).

\begin{figure}
\includegraphics[width=84mm]{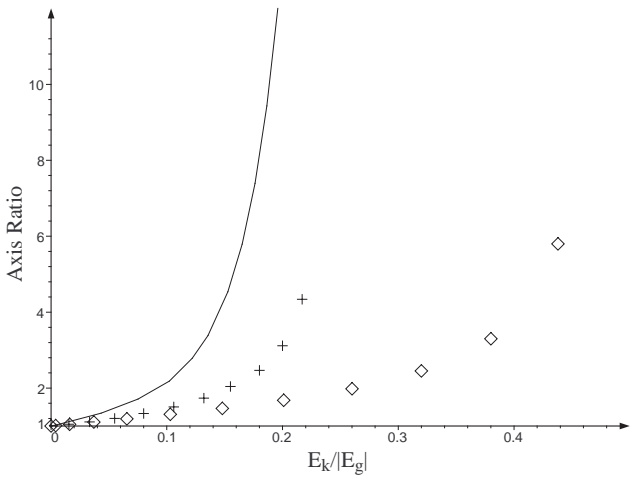}
\caption{Axis ratio versus $E_k/|E_g|$. We see that our formula ($+, \diamond$) underestimates axis ratio.
Choose of $\Delta C$
satisfying virial theorem ($+$) improves situation a bit. Solid line again is result of \citep{eriguchi}.
}
\label{axisratio_vs_TW}
\end{figure}

\begin{figure}
\includegraphics[width=84mm]{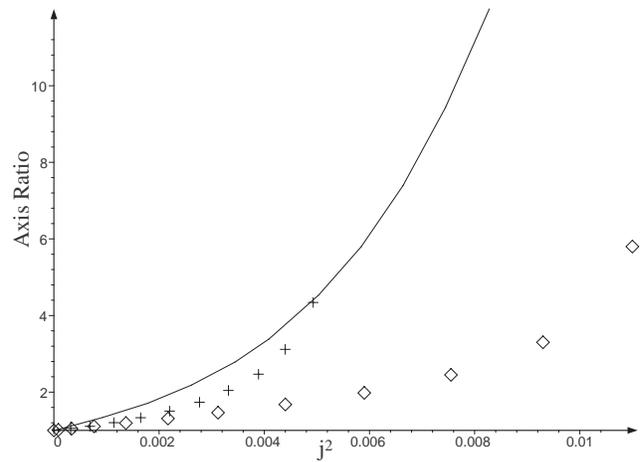}
\caption{Axis ratio vs $j^2$. Everything the same as in Fig.~\ref{axisratio_vs_TW}.
$\Delta C$ from eq.~(\ref{z}) ($+$) gives better approximation to axis ratio compared to
formula (\ref{meanC}) ($\diamond$).
}
\label{axisratio_vs_j2}
\end{figure}

This subsection clearly show importance of constant value $\Delta C$. Best results
are produced with $\Delta C$ from eq.~(\ref{deltaVT}), therefore this value will be used
in the next subsections to investigate influence of differential rotation parameter $A$ and type of rotation law
on formula accuracy.

\begin{table*}
\begin{tabular}{ccclrcl}
\hline
$\Omega_0$&Axis~Ratio&$j^2$& $\frac{E_k}{|E_g|}$  & $\frac{E_g+E_k+U}{E_0}$ & Virial test  $|Z|$& $\Delta C$ \\
\hline \\
$0.25  $&$ 1.01     $&$  8.30 \times 	10^{-5} $&$ 0.004 $&$ -2.05  \times 10^{-6}  $&$ 9 \times 10^{-5}   $&$ 0.004 $\\
$0.50  $&$ 1.05     $&$  3.22 \times 	10^{-4} $&$ 0.02   $&$ -0.98  \times 10^{-5}  $&$ 3 \times 10^{-4}   $&$ 0.02 $\\
$0.75  $&$ 1.11     $&$  6.86 \times 	10^{-4} $&$ 0.03   $&$ -2.10  \times 10^{-5}  $&$ 3\times 10^{-5}    $&$ 0.04 $\\
$1.00  $&$ 1.20     $&$  1.14 \times 	10^{-3} $&$ 0.06   $&$ -3.50  \times 10^{-5}  $&$ 4\times 10^{-5}    $&$ 0.08 $\\
$1.25  $&$ 1.33     $&$  1.65 \times 	10^{-3} $&$ 0.08   $&$ -5.14  \times 10^{-5}  $&$ 6\times 10^{-4}    $&$ 0.13 $\\
$1.50  $&$ 1.51     $&$  2.20 \times 	10^{-3} $&$ 0.11   $&$ -6.84  \times 10^{-5}  $&$ 4\times 10^{-4}    $&$ 0.19 $\\
$1.75  $&$ 1.74     $&$  2.77 \times 	10^{-3} $&$ 0.13   $&$ -8.55  \times 10^{-5}  $&$ 3\times 10^{-5}    $&$ 0.27 $\\
$2.00  $&$ 2.04     $&$  3.32 \times 	10^{-3} $&$ 0.16   $&$ -1.00  \times 10^{-4}  $&$ 1\times 10^{-3}    $&$ 0.37 $\\
$2.25  $&$ 2.47     $&$  3.89 \times 	10^{-3} $&$ 0.18   $&$ -1.18  \times 10^{-4}  $&$ 9\times 10^{-4}    $&$ 0.48 $\\
$2.50  $&$ 3.12     $&$  4.40 \times 	10^{-3} $&$ 0.20   $&$ -1.33  \times 10^{-4}  $&$ 3\times 10^{-4}    $&$ 0.62 $\\
$2.75  $&$ 4.34     $&$  4.93 \times 	10^{-2} $&$ 0.22   $&$ -1.50  \times 10^{-4}  $&$ 1\times 10^{-6}    $&$ 0.77 $\\
\hline
\end{tabular}
\caption{
The same model as in Table \ref{A02jconstMS}, but now $\Delta C$ is derived
numerically from eq.~(\ref{deltaVT}). Virial test show accuracy of solution to eq.~(\ref{deltaVT}).
Comparison with Table \ref{A02jconstMS} and Table 1b of \citet{eriguchi} shows
significant improvement of the total energy. Axis ratio is also closer to results of numerical calculations,
and stability indicator $E_k/|E_g|$ isn't unreasonably high. See also Figs \ref{TWvsj2}--\ref{axisratio_vs_j2}.
}
\label{j-const02}
\end{table*}

}

\subsection{Effects of differential rotation}
{

\begin{figure*}
\includegraphics[width=\textwidth]{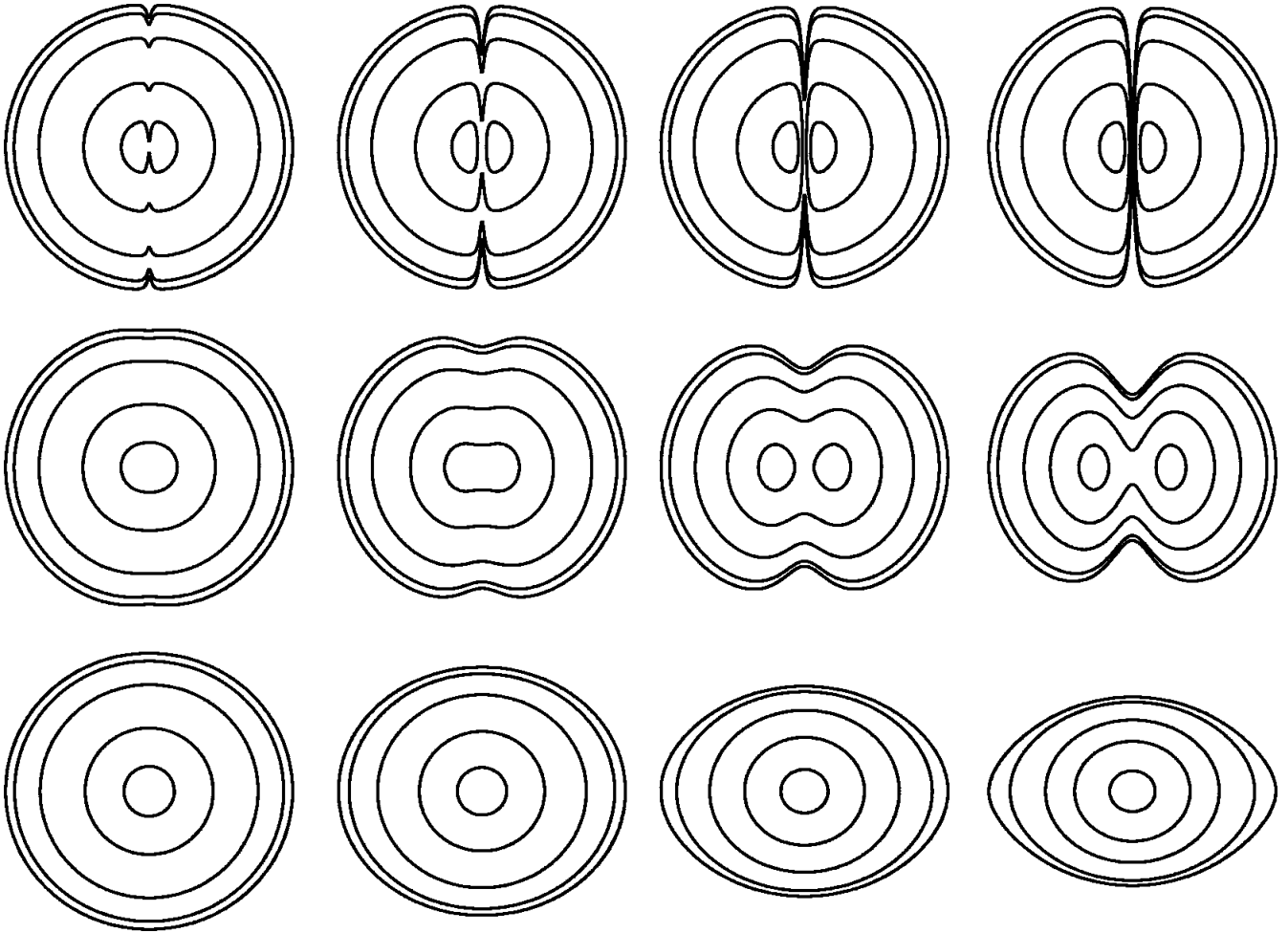}
\caption{Examples of the density distributions given by our formula (\ref{n32_case}).
Results for polytropic model
with $\gamma=5/3$ and so-called $j$-const \citep{eriguchi} rotation law
$\Omega(r)=\Omega_0/(1+r^2/A^2)$ are presented.
Upper row corresponds to diifferential rotation with $A=0.02 R_0$. The value of $\Omega_0$ increases from left to
the right.
Lower row shows behaviour of the almost rigidly rotating star with $A=2 R_0$. For the middle row
$A = 0.2 R_0$ is adopted. $R_0$ is the radius of non-rotating star.
Values of the $\Omega_0$ are, from left: upper row 75, 150, 200, 250; middle row 0.5, 1.0, 1.5, 2.0;
bottom row 0.01, 0.02, 0.03, 0.035.
Constant $\Delta C$ is calculated from eq.~(\ref{deltaC3}).
}
\label{j-const}
\end{figure*}

In addition to the results from previous subsection ($j$-const with $A=0.2R_0$) we have
calculated properties of the almost rigidly ($A=2R_0$) and extremely
differentially ($A=0.02R_0$)  rotating model with the same rotation law.

In all three cases we are able to find value of $\Delta C$ satisfying eq.~(\ref{deltaVT}).
However, this is not enough to find correct solution, because other parameters describing rotating
body may be wrong. This is clearly shown on Fig.~\ref{differ}, where $E_k/E_g$ versus $j^2$
(\ref{j_squared}) is plotted for three cases of differential rotation. Apparent discrepancy for $A=2R_0$
exists. Both $j$-const and $v$-const angular velocity profiles behaves as rigid rotation in this case.
Thus we conclude that our formula is unable to predict correct structure in case of uniform rotation even
if rotation is small.

If rotation is concentrated near rotation axis, like in $A=0.02 R_0$ case, our and numerical results are
of the same order of magnitude.
Quantitative agreement is achieved only for very small values of $\Omega_0$. Let's note that
in this case $\Delta C$ required by virial theorem (\ref{deltaVT}) is slightly below zero (Table~\ref{j-const002}).
This example shows, that $\Delta C$ may also be negative.
All three cases are summarized on Fig.~\ref{differ}.

\begin{table*}
\begin{tabular}{lcccccc}
\hline
$\Omega_0$&Axis~Ratio&$j^2$& $\frac{E_k}{|E_g|}$  & $\frac{E_g+E_k+U}{E_0}$ & Virial test  $|Z|$& $\Delta C$ \\
\hline \\
$0.01  $&$1.04$&$6.10 \times 10^{-6}  $&$1.9 \times 10^{-4}$&$-1.83 \times 10^{-7}$&$9 \times 10^{-5}$&$0.01$\\
$0.02  $&$1.20$&$2.96 \times 10^{-5}  $&$8.7 \times 10^{-4}$&$-8.80 \times 10^{-7}$&$1 \times 10^{-4}$&$0.06$\\
$0.03  $&$1.84$&$1.26 \times 10^{-4}  $&$2.9 \times 10^{-3}$&$-3.58 \times 10^{-6}$&$5 \times 10^{-4}$&$0.15$\\
$0.035$&$2.30$&$4.49 \times 10^{-4}  $&$7.0 \times 10^{-3}$&$-1.13 \times 10^{-5}$&$3 \times 10^{-5}$&$0.24$\\
\hline
\end{tabular}
\caption{Properties of our approximate sequence in case of $j$-const rotation law with $A=2 R_0$,
i.e. almost uniform rotation. In spite of fact that virial test is fulfilled with accuracy of order $10^{-4}$,
comparison of data in this table with numerical results (cf.~Fig.~\ref{differ}) clearly shows that our formula fails
in case of rigid rotation.}
\label{j-const2}
\end{table*}

\begin{table*}
\begin{tabular}{ccclrcr}
\hline
$\Omega_0$&Axis~Ratio&$j^2$& $\frac{E_k}{|E_g|}$  & $\frac{E_g+E_k+U}{E_0}$ & Virial test  $|Z|$& $\Delta C$ \\
\hline \\
$25$&$1.01$&$1.45\times  10^{-4}$&$0.02$&$-0.44 \times10^{-5}$&$4 \times 10^{-4}$&$-0.01$\\
$50$&$1.05$&$4.73 \times 10^{-4}$&$0.07$&$-1.42 \times10^{-5}$&$3\times 10^{-4}$&$-0.02$\\
$75$&$1.12$&$8.31 \times 10^{-4}$&$0.12$&$-2.47 \times10^{-5}$&$4\times 10^{-4}$&$-0.03$\\
$100$&$1.23$&$1.17 \times 10^{-3}$&$0.16$&$-3.36 \times10^{-5}$&$2\times 10^{-4}$&$-0.02$\\
$150$&$1.55$&$1.63 \times 10^{-3}$&$0.23$&$-4.65 \times10^{-5}$&$4\times 10^{-4}$&$0.07$\\
$200$&$2.06$&$1.96 \times 10^{-3}$&$0.28$&$-5.50 \times10^{-5}$&$3\times 10^{-4}$&$0.24$\\
$250$&$2.95$&$2.21 \times 10^{-3}$&$0.32$&$-6.12 \times10^{-5}$&$1\times 10^{-4}$&$0.49$\\
$300$&$5.76$&$2.42 \times 10^{-3}$&$0.34$&$-6.64 \times10^{-5}$&$3\times 10^{-4}$&$0.83$\\
\hline
\end{tabular}
\caption{Properties of sequence with $j$-const rotation law for $A=0.02 R_0$.}
\label{j-const002}
\end{table*}

\begin{figure}
\includegraphics[height=84mm,angle=270]{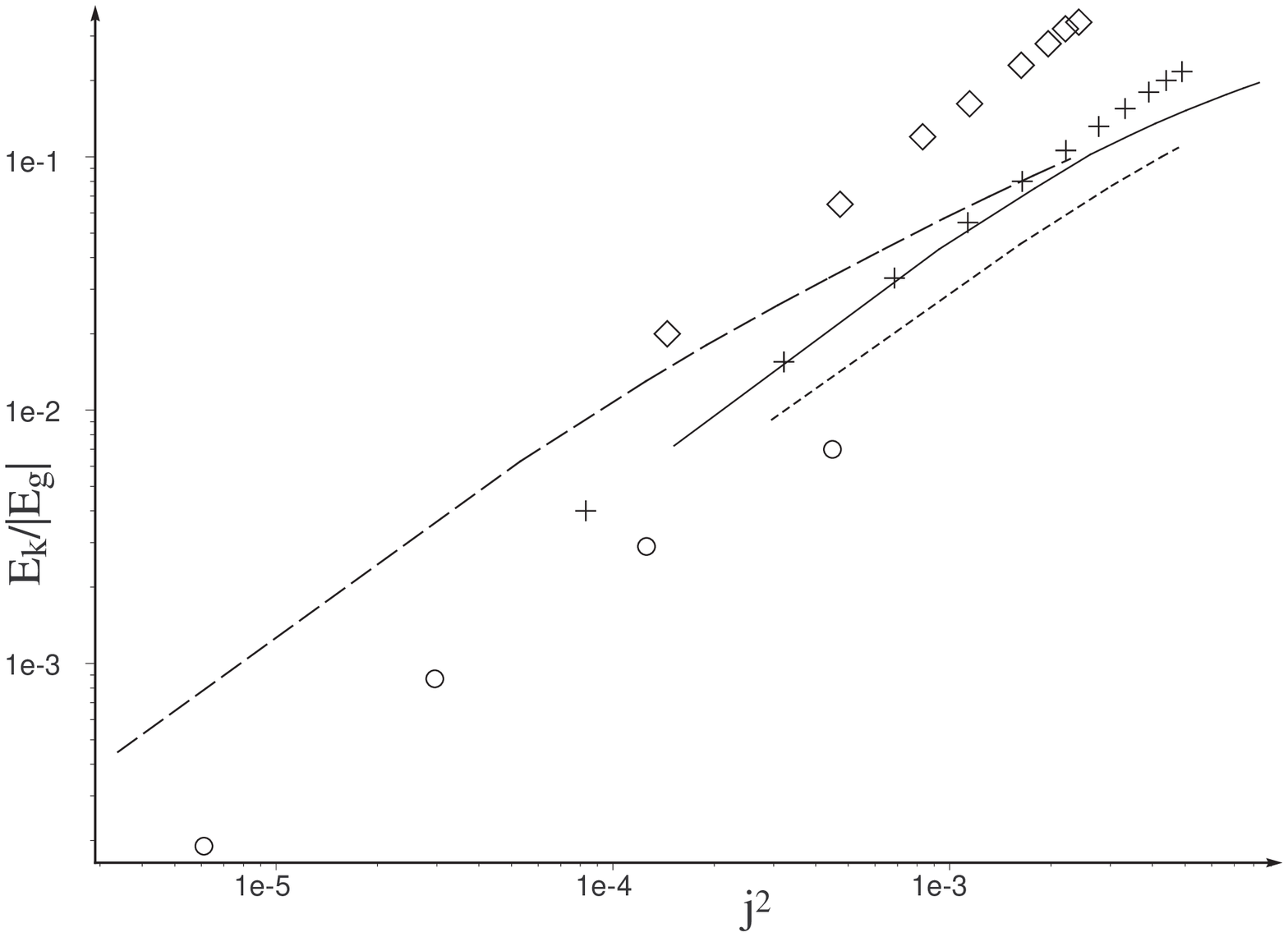}
\caption{Stability indicator $E_k/E_g$ vs $j^2$ for $j$-const angular velocity ($\log-\log$ plot)
for three values of $A$.
Quantitative agreement between our formula (symbols) and numerical results
 (\citealt{eriguchi}, lines)  is achieved for $A=0.2 R_0$ if $E_k/E_g \leq 0.1$. This case is presented as solid line and crosses.
In case of $A = 0.02 R_0$ we have results of the same order, but
they are identical only in case where rotation strength is very small. This case is presented by
dashed line and diamonds. Formula fails  (dotted line and circles) in case of rigid rotation.
}
\label{differ}
\end{figure}
}

Results from this section show, that our formula is able to find correct structure of rotating body
for differential rotation only. Range of application vary with differential rotation parameters, and best results are
obtained in middle range i.e. $A=0.2 R_0$. With extremal case ($A=0.02 R_0$)
quality of our results is significantly degraded.

In next subsection we examine, if this statement depends on rotation law.

\subsection{Rotation law effects}
{
In addition to previously described cases, we have calculated global properties of our model
in case of $v$-const angular velocity profile, with parameter $A=0.2 R_0$ (Table~\ref{v-const02})
and $A=0.02 R_0$ (Table~\ref{v-const002}). Results with  $A=2 R_0$ aren't presented, because they are similar to
$j$-const case (cf.~Table~\ref{j-const2}), where both functions $\Omega(r)$ behave as uniform rotation,
and our formula fails in this case.

Figures \ref{v-const_1} and \ref{v-const_2} show very good agreement of of the global physical
quantities ($E_k/E_g, j^2, E_{tot}$) with numerical results for entire range of
rotation strength covered by both methods. The most extreme case ($A=0.02$)
also behaves well. Axis ratio (Fig.~\ref{v-const_3}) however, clearly distinguish between approximation
and precise solution. Results are quantitatively correct only for small rotation parameters, e.g $j^2 \ll 0.005$
i.e. $E_k/E_g \ll 0.1$.

\begin{table*}
\begin{tabular}{ccclrcc}
\hline
$\Omega_0$&Axis~Ratio&$j^2$& $\frac{E_k}{|E_g|}$  & $\frac{E_g+E_k+U}{E_0}$ & Virial test  $|Z|$& $\Delta C$ \\
\hline \\
$0.25$&$1.05$&$3.29\times 10^{-4}$&$0.01$&$-0.99\times 10^{-5}$&$9\times 10^{-5}$&$0.01$\\
$0.50$&$1.21$&$1.41\times 10^{-3}$&$0.05$&$-4.26\times 10^{-5}$&$2\times 10^{-4}$&$0.05$\\
$0.75$&$1.63$&$3.80\times 10^{-3}$&$0.11$&$-1.12\times 10^{-4}$&$2\times 10^{-4}$&$0.13$\\
$1.00$&$2.31$&$9.22\times 10^{-3}$&$0.21$&$-2.52\times 10^{-4}$&$3\times 10^{-4}$&$0.27$\\
$1.25$&$5.26$&$1.44\times 10^{-2}$&$0.31$&$-3.87\times 10^{-4}$&$5\times 10^{-4}$&$0.47$\\
\hline
\end{tabular}
\caption{Properties of our model with $v$-const rotation law and $A=0.2 R_0$.}
\label{v-const02}
\end{table*}

\begin{figure}
\includegraphics[height=84mm,angle=270]{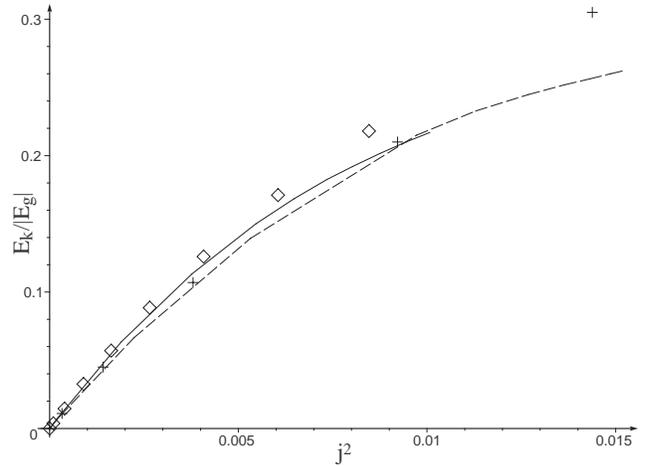}
\caption{$E_k/E_g(j^2)$ for $v$-const rotation law with $A=0.2 R_0$ (dashed, cross) and $A=0.02R_0$ (solid, diamond),
where lines refers to \citet{eriguchi} and symbols refers to our formula with $\Delta C$ from (\ref{deltaVT}).
In this case we have good quantitative agreement with numerical results in both cases up to
$E_k/E_g \simeq 0.1$.}
\label{v-const_1}
\end{figure}

\begin{figure*}
\includegraphics[width=\textwidth]{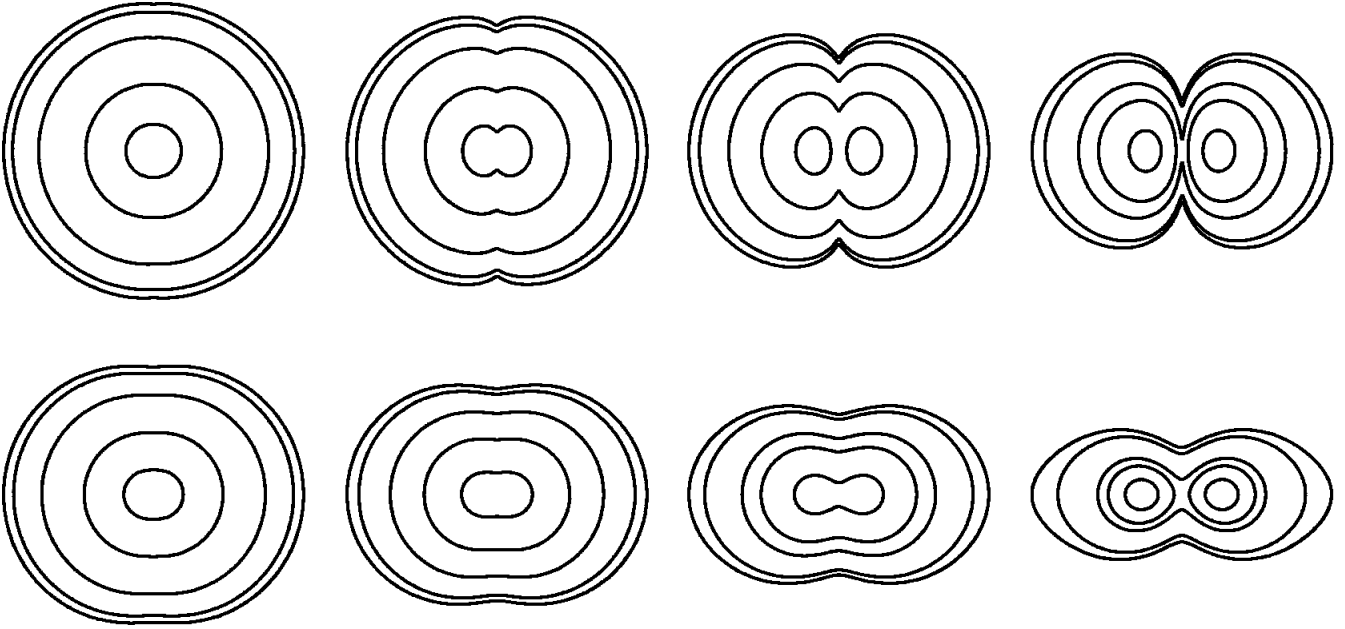}
\caption{Another example of density distributions given by our formula (\ref{n32_case}).
The same as in Fig.~\ref{j-const} for so-called $v$-const rotation law \citep{eriguchi} $\Omega(r)=\Omega_0/(1+r/A)$.
Values of $\Omega_0$ are, from left: in the upper  ($A=0.02R_0$) row: 1.0, 3.0 , 5.0, 7.0;
in the bottom  ($A=0.2R_0$) row: 0.5, 0.75, 1.0 and 1.25.}
\label{v-const}
\end{figure*}

\begin{table*}
\begin{tabular}{ccclrcc}
\hline
$\Omega_0$&Axis~Ratio&$j^2$& $\frac{E_k}{|E_g|}$  & $\frac{E_g+E_k+U}{E_0}$ & Virial test  $|Z|$& $\Delta C$ \\
\hline \\
$1.0 $&$1.02 $&$ 9.81\times10^{-5} $&$ 0.004$&$-0.30\times10^{-5}$&$7 \times 10^{-5}$&$0.01$\\
$2.0 $&$1.09 $&$ 3.94\times10^{-4} $&$ 0.015 $&$-1.20\times10^{-5}$&$1\times   10^{-4}$&$$0.04\\
$3.0 $&$1.21 $&$ 8.96\times10^{-4} $&$ 0.03 $&$-2.75\times10^{-5}$&$7\times   10^{-4}$&$$0.09\\
$4.0 $&$1.41 $&$ 1.63\times10^{-3} $&$ 0.06 $&$-5.00\times10^{-5}$&$5\times   10^{-4}$&$$0.17\\
$5.0 $&$1.73 $&$ 2.65\times10^{-3} $&$ 0.09 $&$-8.09\times10^{-5}$&$1\times   10^{-4}$&$$0.27\\
$6.0 $&$2.24 $&$ 4.08\times10^{-3} $&$ 0.13 $&$-1.23\times10^{-4}$&$3\times   10^{-4}$&$$0.39\\
$7.0 $&$3.03 $&$ 6.05\times10^{-3} $&$ 0.17 $&$-1.77\times10^{-4}$&$2\times   10^{-4}$&$$0.55\\
$8.0 $&$4.60 $&$ 8.46\times10^{-3} $&$ 0.22 $&$-2.38\times10^{-4}$&$3\times   10^{-4}$&$$0.75\\
\hline
\end{tabular}
\caption{Properties of $v$-const sequence for $A=0.02 R_0$.}
\label{v-const002}
\end{table*}

\begin{figure}
\includegraphics[height=84mm,angle=270]{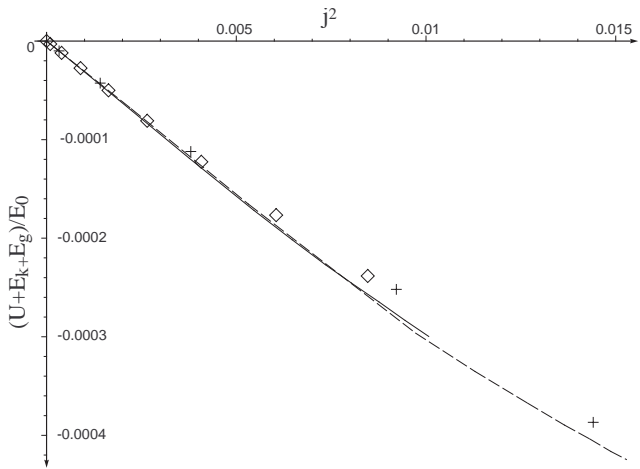}
\caption{$E_{tot} $ versus square of dimensionless angular momentum  $j^2$. Symbols
description is the same as on previous figure, Fig.~\ref{v-const_1}.
}
\label{v-const_2}
\end{figure}

\begin{figure}
\includegraphics[height=84mm,angle=270]{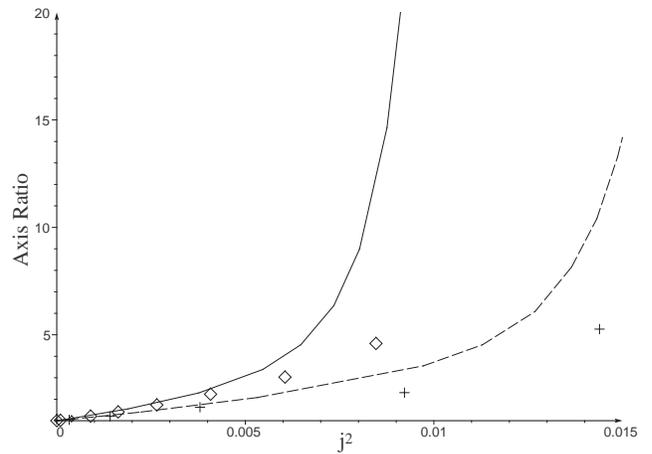}
\caption{Axis ratio versus $j^2$ for $v$-const rotation law. Symbols
description is the same as on previous figures, Fig.~\ref{v-const_1} and Fig.~\ref{v-const_2}.
}

\label{v-const_3}
\end{figure}

}

\section{Discussion\& Conclusions}

Comparison of the results obtained with our approximation formula
(Fig.~\ref{v-const} -- Fig.~\ref{v-const_3}, Table \ref{j-const02}--\ref{v-const02}) with other (\citealt{eriguchi}, Fig.~2--5, Fig.~9,
Table 1 and 2)
shows a correct qualitative behaviour for even the most simplified
version of our approximation formula for a wide range of parameters describing
differential rotation and strength of  rotation.
This make our formula excellent tool for those who are interested in the
structure of barotropic, differentially rotating stars, but do not need exact,
high precision results. It can be applied for qualitative analysis
of structure of rapidly rotating stellar cores (e.g. `cusp' formation,
degree of flattening, off-centre maximum density)
with arbitrary rotation law, also for initial guess for
numerical algorithms. It can also be used as an alternative
for high-quality numerical results for use in a more convenient form
as long as $E_k/|E_g| \ll 0.1$ and we are interested mainly in global properties
of differentially rotating objects.\\

\noindent
\textbf{ACKNOWLEDGMENTS:}

I would like to thank Prof.~K.~Grotowski and M.~Misiaszek for thorough
discussion of the problem, and Prof. M.~Kutschera for critical reading
of the previous version of this article.\\

\label{lastpage}

\end{document}